\documentclass[aps,prd,a4paper,superscriptaddress,nofootinbib,showpacs,twocolumn,showkeys,amsfonts,amssymb,amsmath]{revtex4}
\usepackage{color}
\usepackage{graphicx}
\usepackage{enumitem}
\setlist{nolistsep}

\begin{document}

\title{Dark Energy and Dark Matter from Yang-Mills Condensate \\
and the Peccei-Quinn mechanism }

\author{Andrea Addazi}
\email{andrea.addazi@lngs.infn.it}
\affiliation{University of  L'Aquila \& INFN L'Aquila, Italy}

\author{Pietro Don\`a}
\email[]{pietro\_dona@fudan.edu.cn}
\affiliation{Department of Physics \& Center for Field Theory and Particle Physics, Fudan University, 200433 Shanghai, China}

\author{Antonino Marcian\`o}
\email[]{marciano@fudan.edu.cn}
\affiliation{Department of Physics \& Center for Field Theory and Particle Physics, Fudan University, 200433 Shanghai, China}

\date{\today}

\begin{abstract}
\noindent 
We analyze a model of cold axion Dark Matter weakly coupled with a dark gluon condensate, reproducing Dark Energy. We first review how to recover the Dark Energy behavior using the functional renormalization group approach, and ground our study on the properties of the effective Lagrangian, to be determined non-perturbatively. Then, within the context of $G_{SM}\times SU(2)_{D}\times U(1)_{PQ}$, we consider YMC interactions with QCD axions. We predict a transfer of Dark Energy density into Dark Matter density in a cosmological time that can be tested in the next generation of experiments dedicated to Dark Energy measures.

\end{abstract}

\maketitle

\section{Introduction}

\noindent
Consolidated observations on Supernova Type I a (SN Ia) have established that the universe is undergoing a phase of accelerated expansion. First evidences were provided in 1998 by two independent teams \cite{Riess:1998cb, Perlmutter:1998np}. Since then analyses exploiting SN Ia data set \cite{Kowalski:2008ez}, combined with cosmic microwave background radiation (CMBR) \cite{Spergel:2003cb, Graham:2005xx, Spergel:2006hy, Komatsu:2008hk} through the WMAP satellite observations and larger scale structure \cite{Cole:2005sx,Tegmark:2006az}, have strongly corroborated this scenario. Despite from an experimental point of view the picture has been deeply clarified, nevertheless any compelling theoretical explanation of the origin of the current acceleration of the universe can be yet advocated (see {\it e.g.} \cite{Capozziello:2003tk, Carroll:2003wy, AT}), and the problem has been dubbed in the literature as ``Dark Energy'' (DE). 

Among many possible descriptions stands the simple hypothesis that DE originates from a Yang-Mills field condensate (YMC). A notable analogy is provided by the Higgs field, but there are important caveats to be considered. The Yang-Mills (YM) field advocated to explain DE does not necessarily match the content of matter of the standard model (SM) of particle physics, and might actually represent a different gauge field matter component. A YMC mechanism was first proposed in \cite{YZ1} to allow a primordial inflationary acceleration of the universe, and was described by the renormalization-group-improvement (RGI) action on a Friedmann-Lema\^itre-Robertson-Walker (FLRW) background. The same idea was later on adapted to the explanation of DE in Refs.~\cite{YZ2, YZ3}, respectively in the perturbative two-loop and three-loop analyses of the effective action of SU(N) YM theory. The strategy deployed in \cite{YZ2, YZ3} of accounting for a non-perturbative expansion of the effective action and of retaining only the lower-loops corrections, makes nevertheless not fully reliable the results obtained for the YMC in a regime, the infra-red one, proper of DE. Nevertheless, the core of this proposal relies on considering quantum corrections encoded in the effective action, which can be cast in terms of an effective running coupling constant $g=g(\Theta)$, as derived within the RGI framework~\cite{YZr3, YZr2, YZr3}. The coupling depends on a contraction of the field-strength tensors that plays the role of order-parameter for the YMC, namely $\Theta\equiv -\frac{1}{2}\, F^a_{\mu \nu} F^{a \,\mu \nu}$, and enters the density Lagrangian  $\mathcal{W} = -\frac{1}{4 g^2(\Theta)} F^a_{\mu \nu} F^{a \,\mu \nu}$. Henceforth, sum over repeated internal indices $a$, which run over the dimensions of the Lie-algebra, will be intended. In $g$ a dependence on the square of the renormalization mass-scale $\kappa$ is also present. The latter only denotes the initial point in the renormalization group flow, and must not be confused with a physical scale.

The most serious technical issue plaguing previous analyses \cite{YZ1, YZ2, YZ3} concerns the stability of the results obtained in the perturbative approach to the computation of the effective action. At higher orders than the three-loops expansion, the appearance of additional terms in the effective action for  SU(N) YM-theories might spoil the DE behavior, which totally relies on an ultraviolet perturbative expansion. Conversely, summarizing the analysis in \cite{Dona:2015xia}, we proceed in Sec.~II and Sec.~III to show that a fully non-perturbative approach is possible. 

First, in Sec.~II we prove that under mild and general assumptions, which are basically the existence of a minimum in $\Theta$ in the non-perturbative effective Lagrangian, a DE behavior is recovered. 

Then in Sec.~III, by making use of the non-perturbative techniques mutated from the  functional renormalization group (FRG) procedure, which is more adequate to be used in the confining infrared limit of the theory, we show that such a minimum indeed exists, at least for the case of SU(2), and we provide the explicit example of the latter. 
We can state general requirements for the effective density Lagrangian $\mathcal{W}\left(\Theta \right)$ that must be fulfilled in order to obtain a YMC model for DE:
\begin{enumerate}
\item[PI)] $\mathcal{W}(\Theta)$ has a non trivial minimum at some energy scale $\Theta_0\approx \Lambda_D^4$; 
\item[PII)]  $\mathcal{W}(\Theta)$ posses a perturbative limit, which resembles the one-loop result derived by Savvidy \cite{Savvidy:1977as};
\item[PIII)] $\mathcal{W}(\Theta)$ shows the UV asymptotic behavior ($\Theta\gg \Lambda_D^4$) 
of being at least linear in $\Theta$, which in turn is linear in the bare Yang-Mills action. 
\end{enumerate}
The effective action $\mathcal{W}(\Theta)$ will be in general equipped with a characteristic energy scale $\Lambda_D$. A YMC then forms whenever the minimum of $\mathcal{W}(\Theta)$ is reached. 

Sec.~IV and Sec.~V are the original parts of this work. In Sec.~IV we discuss the dark YMC model in the context of a minimal Standard Model (SM) extension $SU(3)_{c}\times SU(2)_{L}\times U(1)_{Y}\times SU(2)_{D} \times U(1)_{PQ}$, where $U(1)_{PQ}$ is the Peccei-Quinn global axial symmetry, spontaneously broken and associated to a QCD axion. 

In Sec.~V we emphasize that while DE is described by the dark YMC, the QCD invisible axion provides a good candidate for cold DM. We retain this minimal extension strongly motivated by the strong CP problem. The whole model we present here has an important difference with respect to traditional QCD axion theories: axions can interact with the dark YMC in a EFT framework. In particular, a part of the DE density can be transferred to the DM density during a cosmological time of $1\div 10\, \rm Gyrs$ or so. We will estimate the rate of this process and its cosmological limits. 

In Sec.~VI, we spell out some conclusions and outlooks.

\section{YMC as Dark Energy}

\noindent
To shed light on the behavior of the YMC, and check whether it can solve the problem of DE, we assume henceforth a flat FLRW universe, the line element of which is cast in terms of comoving coordinates, {\it i.e.} $ds^2=dt^2-a^2(t)\delta_{ij}dx^idx^j$, with $t$ cosmological time. In the simplest case of a universe filled only with the YMC minimally coupled to gravity, the effective action reads
 \begin{eqnarray}
 \mathcal{S}=\int \sqrt{-{g}}~\left[-\frac{\mathcal R}{16\pi G}+\mathcal{W}(\Theta)
 \right] ~d^{4}x,
 \label{S}
 \end{eqnarray}
with ${g}$ the determinant of the metric $g_{\mu\nu}$, and $\mathcal{R}$ the scalar Ricci curvature. By variation of $\mathcal{S}$ with respect to the metric $g^{\mu\nu}$, one obtains the Einstein equation $G_{\mu\nu}=8\pi G\, T_{\mu\nu}$, the energy-momentum tensor of the YMC being
 \begin{eqnarray}\label{T_munu}
 \!\!\!\!\! T^{\mu\nu}\!=\!\sum_{a=1}^{3}~ {}^{(a)}T^{\mu\nu}=\sum_{a=1}^{3}~ g^{\mu\nu}  \mathcal{W}\left(\Theta\right) - 2 \dfrac{\partial \mathcal{W}}{\partial \Theta} F_a^{\gamma\mu}F^{a}{}_\gamma^{\phantom{\gamma}\nu}.
 \end{eqnarray}
 The YM tensor can be cast in terms of the structure constants $f^{abc}$ of the SU(N) gauge-group under scrutiny, and generally reads $F^{a}_{\mu\nu}=\partial_{\mu}A_{\nu}^a-\partial_{\nu}A_{\mu}^a+f^{abc}A_{\mu}^{b}A_{\nu}^{c}$. From now on, we focus on the SU$(2)$ gauge-group, for which $f^{abc}=\epsilon^{abc}$. We then pick out a gauge that preserves isotropy and homogeneity of the FLRW background, by assuming gauge fields to be functions only of the cosmological time $t$, and choosing their components $A_0=0$ and $A_i^a=\delta_i^aA(t)$. The YM tensor then assumes a simplified form, and its non vanishing components read $F^{0a}_{i}= E/3$. The order parameter $\Theta$ can be cast in a  simple form, {\it i.e.} $\Theta=E^2$, and the energy-momentum tensor is found to be isotropic, with energy and pressure densities given by
\begin{eqnarray} \label{rho_p}
\!\!\!\!\!\!\rho_\text{YMC}\!=\!-\mathcal{W}(\Theta)\!+\! 2\mathcal{W}'(\Theta)\Theta, \\  p_{\text{YMC}}\!=\!\mathcal{W}(\Theta)\!-\! \frac{2}{3}\mathcal{W}'(\Theta)\Theta .\,
\end{eqnarray}
Consequently, the equation of state (EOS) of the YMC is immediately recovered to be
 \begin{eqnarray}\label{eos}
 w_\text{YMC}\equiv\frac{p_\text{YMC}}{\rho_\text{YMC}}=-\frac{\mathcal{W}-\frac{2}{3}\mathcal{W}' \, \Theta}{ \mathcal{W} -2 \mathcal{W}' \, \Theta} = -\frac{1-\frac{2}{3}\frac{\mathcal{W}'}{\mathcal{W}} \, \Theta}{1 -2 \frac{\mathcal{W}'}{\mathcal{W}}  \, \Theta} \, .
 \end{eqnarray}
If we require the YM theory to condensate (property PI in Sec.~I), then the function $\mathcal{W}(\Theta)$ must have a non trivial minimum, which implies that $\mathcal{W}'$ vanishes at some $\Theta_0$. At $\Theta_0$, the YMC has an EOS proper of the cosmological constant, since $w_\text{YMC}=-1$. 

In the high-energy-scale regime $\Theta\gg \Lambda_D^4$, as a consequence of property PIII we recover for the condensate an EOS of radiation for the YMC, {\it i.e.} characterized by $w_\text{YMC}=1/3$, in analogy with the perturbative analysis \cite{YZ1,YZ2,YZ3}.

We may generalize this analysis, and resort to a description of the universe that takes into account YMC, matter and radiation, treated in terms of their EOS. Since we have assumed {\it ab initio} the universe to be flat, fraction densities must sum up to the identity, {\it i.e.} $\Omega_\text{YMC}+\Omega_m+\Omega_r=1$, with fractional energy densities defined as $\Omega_\text{YMC}\equiv \rho_\text{YMC}/\rho_{tot}$,   $\Omega_{m}\equiv \rho_m/\rho_{tot}$,  $\Omega_{r}\equiv \rho_r/\rho_{tot}$, and total energy density $\rho_{tot}\equiv\rho_\text{YMC}+\rho_m+\rho_r$. Friedmann equations then read
\begin{eqnarray}\label{friedmann1}
 &&
 \left(\frac{\dot{a}}{a}\right)^2=\frac{8\pi G}{3}(\rho_\text{YMC}+\rho_m+\rho_r),\\ \label{friedmann2}
&&
\frac{\ddot{a}}{a}=-\frac{4\pi G}{3}(\rho_\text{YMC}+3p_\text{YMC}+\rho_m+\rho_r+3p_r),
\end{eqnarray}
with \textit{dot} denoting time-derivative. If we assume no interaction between the three energy components, the dynamical evolution is dictated by energy conservation:
 \begin{eqnarray}\label{aa1}
 &&\dot{\rho}_\text{YMC}+3\frac{\dot{a}}{a}(\rho_\text{YMC}+p_\text{YMC})=0,\\
 &&\dot{\rho}_m+3\frac{\dot{a}}{a}\rho_m=0, \qquad \dot{\rho}_r+3\frac{\dot{a}}{a}(\rho_r+p_r)=0. \label{aa3}
 \end{eqnarray}
From Eqs. \eqref{aa3}, the standard evolutions of the matter and radiation components easily follow, {\it i.e.} $\rho_m\propto a^{-3}$ and $\rho_r\propto a^{-4}$, while less obvious turns out to be the evolution of the YMC. Inserting \eqref{rho_p} into \eqref{aa1} yields
 \begin{eqnarray}\label{y_evolution_eq}
\dot{\Theta} \left( \mathcal{W}' + 2 \mathcal{W}''\, \Theta \right) + 4 \frac{\dot{a}}{a} \mathcal{W}'\, \Theta =0\,,
 \end{eqnarray}
a quite compact form that is integrable for any regular enough $\mathcal{W}$. The result is then easily derived: 
 \begin{eqnarray}\label{y_evolution}
\sqrt{\Theta} \mathcal{W}'\left(\Theta\right) = \alpha a^{-2}\,,
 \end{eqnarray}
where $\alpha$ is a coefficient of proportionality that depends on the initial conditions, to be fine-tuned in order to recover the redshift $z$ at which the universe transits into its dark energy phase. At very high redshift, in the limit $\Theta \gg \Lambda_D^4$, Eq. \eqref{y_evolution} entails an increase of the order-parameter $\Theta$. Then Eq. \eqref{eos} encodes the EOS' parameter $w_\text{YMC}\rightarrow 1/3$, and the YMC starts behaving as a radiation component, as expected from asymptotic freedom at high energy. At small redshift, the expansion of the universe requires the LHS of Eq. \eqref{y_evolution} to asymptotically vanish. This occurs for the extremal value of $\Theta_0$, at which the EOS' parameter converges towards $w_\text{YMC}=-1$, which implies a DE behavior.

\section{FRG and YMC in the SU(2) extra sector}
\label{secYMC}
\noindent
In the previous section we reviewed the consequences of properties PI-III for the cosmological evolution of the YMC. We shall give hints now that the YMC so far discussed indeed exists, beyond the perturbative approximation \cite{YZ1,YZ2,YZ3}. 

The FRG approach, a tool developed to study the non-perturbative flow of a QFT, will provide us with the tools necessary to achieve this purpose. The scale-dependence of the flowing action is recovered by solving the FRG Equation (FRGE) \cite{Wetterich:1993yh, Morris:1993qb}:
\begin{equation}
\label{WetterichEq}
\partial_{k}\Gamma_{k}=\frac{1}{2}\mathrm{STr}\left(\Gamma_{k}^{(2)}+R_{k}\right)^{-1}\partial_{k}R_{k},
\end{equation}
in which $k$ is a (running) cutoff scale that allows us to interpolate smoothly between the microscopic action $\Gamma_{k \rightarrow \infty}$ and the full quantum effective action $\Gamma_{k \rightarrow 0}$; the Super-Trace $\rm STr$ is over all discrete indexes and fields, and encodes summation over the eigenvalues of the Laplacian; the quantity $R_k$ denotes a mass-like regulator function that suppresses quantum fluctuations with momenta lower than an IR momentum-cutoff-scale $k$, implementing the Wilsonian renormalization group flow idea with a momentum-shell wise integration of the path-integral.

The FRGE has been extensively applied to SU(N) YM-theories \cite{Fischer:2002hn,Fischer:2006vf,Fischer:2008uz,FRG,Pawlowski:2003hq}, and to the study of YMC --- see {\it e.g.} \cite{Reuter:1994zn,Reuter:1997gx,Gies:2002af} and \cite{Eichhorn:2010zc} (the most recent work).

It is impossible to solve \eqref{WetterichEq} exactly, unless we deploy minimal approximations. We might make an {\it ansatz} on the functional form of the effective action to solve \eqref{WetterichEq}, but this would shift away from our purpose of determining $\mathcal{W}(\Theta)$. 

We shall instead proceed to reconstruct $\mathcal{W}(\Theta)$, taking into account the energy-flow generated by the propagator of the bare action, rather than from the propagator of the full (unknown) effective action. In perturbation theory this is equivalent to working at one loop, so we expect to reproduce a result that is similar to one in \cite{Savvidy:1977as}. A similar analytical calculation in the full flow equation is beyond currently available techniques. To perform a better analysis, some form of interpolated propagator from the UV (free) regime to the IR (interactive) regime should be deployed \cite{Eichhorn:2010zc}. This would amount to use numerical methods and an effective description. Although conclusions on the DE behavior will be quantitatively affected by this approximation, we believe that these arguments strengthen the existence of the minimum in the effective action, and thus the appearance of the DE phase. 

The computation is performed in the framework of the background field method. The two key ingredients that enter into \eqref{WetterichEq} are a wise choice of the YM background field (following \cite{Eichhorn:2010zc} we will use a self-dual background to avoid unnecessary complication with negative eigenvalues of the laplacian) and the deployment of the simplest possible regulator function (mass-like cutoff $R_{k}\left(\mathcal{D}\right)=k^2$ in all the sectors of the trace). The resulting effective Lagrangian (after identifying the characteristic energy scale with $\Lambda_D$) is the following:
\begin{eqnarray}
\label{eqLeff}
\!\!\!\!\!\!\!\!\!\!\! \mathcal{W}(\Theta)=
\frac{g^2 \Theta}{2 \pi^2}\! \int_0^\infty \!\!\frac{ds}{s}e^{-s\sqrt{\frac{\Lambda_D^4}{g^2 \Theta}}} \!\!\left(\!\frac{1}{4\sinh^2\left(s\right)}\!+\!1\! -\! \frac{1}{4s^2}\!\right)\!\!,
\end{eqnarray}
where the coupling constant $g$ appearing in the equation is the bare coupling constant we used as initial condition for the integration of the RG flow. For a detailed derivation of \eqref{eqLeff} we refer to \cite{Dona:2015xia}.

\begin{figure}[!h]
\label{picLeff}
\includegraphics[scale=0.80]{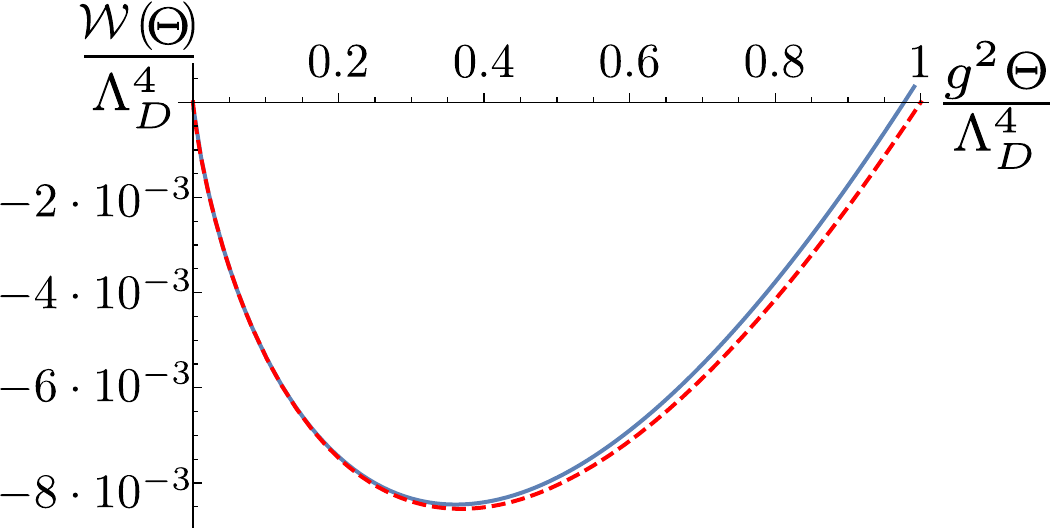}
\caption{Plot of the function \eqref{eqLeff} (blue-continuous) and the one-loop \cite{Savvidy:1977as} (red-dashed). Notice the presence of a non zero global minimum for $\frac{g^2 \Theta_0}{\Lambda_D^{4}} \approx 0.361$.}
\end{figure}

From Fig.1 it is evident that the function \eqref{eqLeff} has a non zero global minimum. The exact position of this minimum can be computed numerically, and in terms of dimensionless quantities is found to be $\frac{g^2\,\Theta_0}{\Lambda_D^{4}} \approx 0.361$, which is consistent with what we expected from the property PI stated in Sec.~I.

Moreover it is possible, as expected, to reproduce the one-loop result derived by Savvidy \cite{Savvidy:1977as} by  computing the asymptotic expansion of $ \mathcal{W}(\Theta)$ for small values of the YM UV coupling constant $g$. In this limit:
\begin{equation}
 \mathcal{W}(\Theta)  \approx  \frac{11}{48 \pi^2} g^2 \Theta \mathrm{Log} \left(\frac{\Lambda_D^4}{g^2 \Theta}\right) \,.
\end{equation}

Finally, we verify that the SU(2) YMC so far discussed evolves from a radiation like component to a DE one. We come back to \eqref{y_evolution}, and estimate the characteristic scale of the condensate $\Lambda_{D}$ by comparing the ``predicted'' YMC fractional energy density at low redshift with the measured DE fractional energy density $\Omega_\Lambda = 0.735$. We then find that for a wide range of initial conditions --- the parameter $\alpha$ in \eqref{y_evolution} --- $\Lambda_{D}\approx 3.2 h^{1/2} 10^{-3} eV$. As noticed in \cite{YZ1,YZ2,YZ3}, this is a very low energy scale compared to typical energy scales in particle physics, thus the SU(2) Yang-Mills interaction must be assumed to describe a dark sector. 

We can study the evolution of the YMC energy density and its EOS for different values of $\alpha$, and still find the same asymptotic values. The value of the cosmological constant and the value of $z$ at the transition epoch to dark energy are known by experimental evidences, and are provided with statistical errors. We can then fine-tune the parameter $\alpha$ to be consistent with experimental data, in the window allowed by current data. Results are summarized in Fig.~2.
\begin{figure}[!h]
\label{pic2}
\includegraphics[scale=0.95]{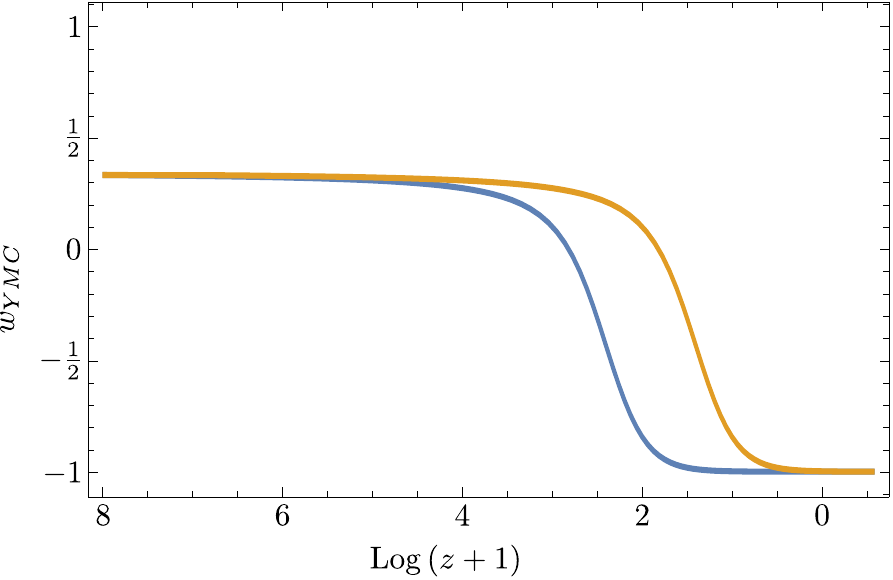}
\caption{The EOS for YMC models of DE that enjoy different initial conditions. The decreasing of the value of the parameter $\alpha$ amounts to a more realistic start of the DE behaviour at values of the redshift closer to zero.} 
\end{figure}

A stability analysis has been performed for our model in the fully interacting model~\cite{Dona:2015xia}. The position of the fixed point can be estimated numerically. There exists a unique fixed point for every positive value of the coupling parameter and is always attractive.

\section{Dark Yang-Mills phase transition}

\noindent
In the previous section, we have studied the first order phase transition of the dark Yang-Mills theory in the framework of the FRG approach. From now on we argue how this model can be extended in order to unveil the origin of DM. We start by considering in this section heuristic arguments in favor of a mechanism of evaporation of the gluon condensate of DE into DM. Then in the next section we focus on the instantiation of DM in our model in terms of an axion field coupled to the invisible Yang Mills sector. \\

In this section we sketch the evaporation of the gluon condensate at finite temperature
from the point of view of glueballs approach.  At low temperature gluons are {\it frozen} (inside the gluon condensate) at a characteristic wavelength $\Lambda_{D}^{-1}$. At temperature higher that a critical temperature $T_{c}\simeq \Lambda_{D}$, gluons have enough kinetic energy to escape from the condensate, and the evaporation process will start. A first order phase-transition happens at $T_{c}$. For $T<T_{c}$ the gluon condensate is dominant, while for $T\!>\!\!>\!T_{c}$ it can be considered as a gas of free gluons. Formation of a condensate breaks conformal symmetry; dilatons are the only degrees of freedom left, and can be described as been governed by the Lagrangian     
\begin{equation*}
\mathcal{L}=\frac{1}{2}(\partial_{\mu}\rho)^{2}-V(\rho),
\end{equation*}
in which $V(\rho)$ has a certain minimum at $\rho=\rho_{0}$. Oscillations around this minimum describe excitations of a scalar glueball, {\it i.e.}
\begin{equation*}
V(\rho)=\frac{1}{2}M_{g}^{2}(\rho-\rho_{0})^{2}+O[(\rho-\rho_{0})^{3}],
\end{equation*}
where $M_{g}$ is related to the second derivative of $V$ in its minimum. 

In the mean field approximation, one can describe 
the pressure of the dilaton field as 
\begin{equation*}
P(\rho,T)=-T\mathcal{I}_{1},
\end{equation*}
in which 
\begin{eqnarray}
\mathcal{I}_{1}=\int \frac{d^{3}p}{(2\pi)^{3}}{\rm ln}[1-e^{-\omega(\rho)/T}],\nonumber \\ 
\omega(\rho)=\sqrt{p^{2}+m^{2}(\rho)}, \qquad m^{2}(\rho)=\frac{\partial^{2}V}{\partial \sigma^{2}}, \nonumber 
\end{eqnarray}
having defined $m^{2}(\rho_{0})=M_{g}^{2}$. 
The thermodynamic potential can be then related to the pressure as
\begin{equation*}
\Omega(\rho,T)= V(\rho)-P(\rho,T).
\end{equation*}
For $T>\!\!>T_{c}$, glueballs do not exist and dilaton's fluctuations cannot be relevant degrees of freedom of the system. However, gluon momenta experience an infrared cutoff at the scale $\Lambda_{D}$, so that the free-gluon gas pressure reads 
\begin{equation*}
P(\rho,T)=-(N_{c}^{2}-1)\,T\,\mathcal{I}_{2}\,,
\end{equation*}
in which
\begin{equation*}
\mathcal{I}_{2}=\int \frac{d^{3}p}{(2\pi)^{3}}{\rm ln}[1-e^{-p /T}]\theta[p-\bar{p}(\sigma)].
\end{equation*}
In the latter expression $\bar{p}(\sigma)$ denotes a cutoff that is a function of the dilaton field vev. Furthermore, it is an appropriate function the asymptotic limits of which are $\bar{p}(\sigma)\rightarrow \infty$ for $\rho \rightarrow \rho_{0}$ and $\bar{p}(\sigma)\rightarrow {\rm const}$ for $\rho\rightarrow 0$ (the cutoff vanishes for $\rho\!<\!\!<\!\rho_{0}$).

As emphasized in the literature, these two regimes are expected to be related by a first order phase transition, {\it i.e.} a discontinuous increase of the dilaton\footnote{We are not interested in developing here a detailed numerical study of these features, which nonetheless are still present in the literature for several YM toy-models and/or realistic models. We only limit ourself to conclude that an ambiguity in the definition of the S-matrix will be naturally solved if the Universe evolves toward a big crunch rather than an infinite expansion. In fact, during the final contraction, the thermal bath will inevitably reach $T>T_{c}$, and the condensate will evaporate \cite{Kolb:1990vq}. However, the introduction of an axion coupled to the dark strong sector can increase the evaporation rate \cite{evaporation}, as we will discuss in the next section. } field profile $\rho(T)$ around $T_{c}$ \cite{GC1,GC2,GC3,Drago:2001gd} .

\section{Dark Gluon condensate and QCD axion condensate}

\noindent 
Within the framework of the minimal model considered in the previous sections, we cannot identify any viable DM candidate. In this section, we discuss how to combine our scenario with the QCD invisible axion paradigm. The latter field is associated to the solution of CP problem, and provides a good candidate for cold and hot DM. Recently a mechanism involving QCD axions for an electroweak scale relaxation was also suggested \cite{Graham:2015cka}. We then suggest to extend the SM encoding an extra $SU(2)_{D}\times U(1)_{PQ}$ gauge sector. 

It is renown that a CP-violating term in QCD $\frac{\theta_{QCD}}{32\pi^{2}}  G_{\mu\nu}^{a}\widetilde{G}^{a\mu\nu}$ can lead to large neutron electric dipole moment. However, measurements of the neutron electric dipole moment actually constrain $\theta_{QCD}<10^{-10}$. Furthermore, it is also common knowledge that the CP-violating term can be shifted away by the ordinary Peccei-Quinn mechanism. 

The complete Lagrangian of the $SU(3)_{c}$ casts as 
\begin{eqnarray*}
&
\mathcal{L}=-\frac{1}{4}G_{\mu\nu}^{a}G^{a\mu\nu}+\frac{\theta_{QCD}}{32\pi^{2}}G_{\mu\nu}^{a}\widetilde{G}^{a\mu\nu}
\nonumber\\&
+\sum_{f}\bar{q}_{f}(i\gamma_{\mu}D_{\mu a}^{b}-m_{r}\delta_{a}^{b})q_{rb}\,,
\end{eqnarray*}
and is invariant under global axial $U(1)_{PQ}$ Peccei-Quinn transformations, when the following shift are implemented   
\begin{equation*}
\theta_{QCD}\rightarrow \theta_{QCD}-{\rm arg}{\rm det} M\,,
\end{equation*}
$M$ denoting the quark mass matrix with eigenvalues $m_{r}$. The shift is equivalent to say that quark mass matrix phases are extra sources of CP violations, {\it i.e.} that the axial $U(1)_{PQ}$ acts on the vacuum as $e^{i\alpha_{r} \bar{Q}_{5}}|\theta\rangle=|\theta+{\rm arg}{\rm det} M\rangle$. The Peccei-Quinn solution of CP problem consists in promoting $\bar{\theta}_{QCD}=\theta_{QCD}+{\rm arg}{\rm det} M$ to a dynamical field $\bar{\theta}_{QCD}=a/f_{a}$, $a$ being the axion field, Goldstone boson of the spontaneously broken global axial symmetry $U(1)_{PQ}$, and the scale $f_{a}$ being the spontaneous symmetry breaking scale of $U(1)_{PQ}$. The latter can be minimally realized through a complex scalar $\sigma=\frac{f_{a}}{\sqrt{2}}e^{ia/f_{a}}$ with a sombrero-like potential. Notice however that the Peccei-Quinn one is not a symmetry of the quantum theory, and $a$ gets an expectation value induced by the QCD sector, which reads 
\begin{equation*}
\left\langle \frac{\partial V_{\rm eff}}{\partial a} \right\rangle \Big|_{\langle a \rangle}=\frac{1}{32\pi^{2}}\frac{1}{f_{a}}\left\langle G_{\mu\nu}^{a}\widetilde{G}^{a\mu\nu} \right\rangle \Big|_{\langle a \rangle}=0.
\end{equation*}
The effective potential has a periodicity $\theta -\frac{\langle a \rangle}{f_{a}}$, so that $a$ is forced to get an expectation value $\langle a \rangle=f'_{a}\theta'$. For instance, the axion effective potential in the dilute gas approximation is $V_{\rm eff}(a)\simeq K\cos (a/f_{a})$, where $K\sim \Lambda_{QCD}^{4}$. Thus the physical axion field turns out to be $\tilde{a}=a-\langle a \rangle$, and the strong sector generates a mass term for $\tilde{a}$ as 
\begin{equation*}
m_{a}^{2}=\left\langle \frac{\partial^{2} V_{\rm eff}}{\partial a^{2}}\right\rangle \Big|_{\langle a \rangle}=\frac{1}{32\pi^{2}}\frac{1}{f_{a}}\frac{\partial}{\partial a} \left\langle G_{\mu\nu}^{a}\tilde{G}^{a\mu\nu} \right\rangle \Big|_{\langle a \rangle},
\end{equation*}
which can be also expressed as 
\begin{equation*}
m_{a}^{2}=\frac{A^{2}}{f_{a}^{2}}\frac{VK}{V+K{\rm Tr}M^{-1}}.
\end{equation*}
In the latter expression $A$ is the color anomaly of the $U(1)_{PQ}$ current, ($A\!=\!1$ for $N_{f}\!=\!N_{c}$), $K$ is related to $V_{\rm eff}$ as mentioned above, $M$ is the quark mass matrix and $V\sim \langle \bar{q} q \rangle \sim \Lambda_{QCD}^{3}$. By virtue of $V$, related to the chiral symmetry breaking, axion gets a small mixing with pseudo-Goldstone bosons  $\pi,\eta$. Using $(m_{u}+m_{d})\langle \bar{q}q\rangle=m_{\pi}^{2} f_{\pi}^{2}$ (neglecting s-quark contribution) we can rewrite the axion mass in the useful form 
\begin{equation*} 
m_{a}=\frac{Az^{1/2}}{1+z}\frac{f_{\pi}}{f_{a}}m_{\pi}\simeq A\left(\frac{10^{6}\, \rm GeV}{f_{a}} \right)\times 6\, \rm eV,
\end{equation*}
with $z=m_{u}/m_{d}\simeq 0.6$ and $m_{\pi},f_{\pi}$ pion mass and decay constant respectively. 

Limits on the masses of QCD invisible axions (KSVZ and DFZX models) are constrained in the range $10^{-1}\div 10\, \rm meV$, while $f_{a}\simeq 10^{9}\div 7\times 10^{10}\, \rm GeV$, placed by  Axion Cold DM production from misalignment mechanism ADMX, CAST, Telescope searches, Globular cluster stars, white dwarfs cooling, SN1987A \cite{PDG} and CMB Constraints \cite{Ejlli:2013uda}.

Within the framework of effective field theory, we can introduce an interaction term of QCD axion with the dark gluon condensate $\mathcal{O}_{aF\widetilde{F}}=\frac{1}{\mathcal{M}}aF_{\mu\nu}\widetilde{F}^{\mu\nu}$. Notice that this additional operator does not spoil the analysis in Sec.~\ref{secYMC}, in fact its only non vanishing contribution to the quadratic part is proportional to $\frac{\delta}{\delta A_\mu}F\widetilde{F}$, which is the first variation of a topological term. In principle a dark $SU(2)_{D}$ sector has a CP violating term. This term induces an extra contribution to the axion mass coming from dark gluon condensate, {\it i.e.}
\begin{equation*}
\Delta m_{a}^{2}=\ \frac{1}{32\pi^{2}}\frac{1}{\mathcal{M}}\frac{\partial}{\partial a} \left\langle F_{\mu\nu}^{a}\tilde{F}^{a\mu\nu} \right\rangle \Big|_{\langle a \rangle}.
\end{equation*}
As a consequence it turns out that $m_{a}\sim \Lambda_{D}^{2}/\mathcal{M}+\sqrt{m_{q}\Lambda_{QCD}^{3}}/f_{a}$. Contrary to the scale $f_{a}$, $\mathcal{M}$ is not directly constrained by coupling with SM particles. However, the extra contribution will be much smaller than the ordinary QCD one, and thus can be neglected: $\Delta m<<\Lambda_{D}\simeq 10^{-4}\, \rm eV$. 

Nonetheless, the interaction term $\mathcal{O}_{aF\widetilde{F}}$ is a gateway from DE to DM. By virtue of this interaction, a dgluon, namely a gluon of the dark sector, can be converted into an axion in the dark condensate. This is a process $gg\rightarrow ga$ mediated by an off-shell dgluon, {\it i.e.} a three-dgluon and a axion-bigluon vertices. 

\begin{figure}[!h]
\label{a}
\includegraphics[scale=0.082]{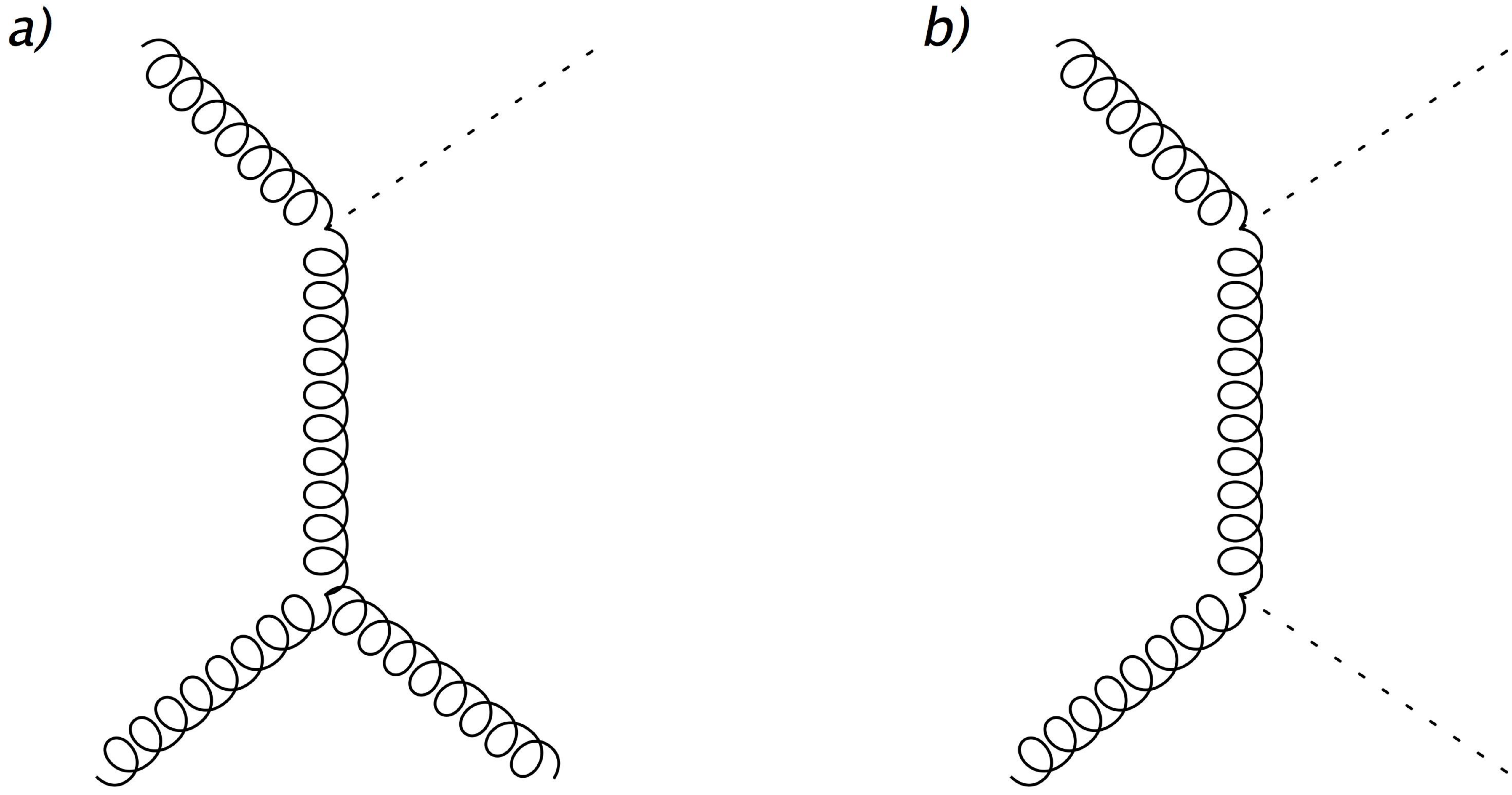}
\caption{Feynman diagrams triggered at tree-level by the Chern-Simons term: curly lines represent dark gluon, while dashed lines represent axions. Decay of DE into DM are originated by dgluon-dgluon scattering, with the exchange of intermediate dgluon quanta, and the production of dgluon and axion particle (Fig.a) or two-axions (Fig.b). The former process is first order in the characteristic energy scale of the theory $\mathcal{M}$, while the latter is suppressed by an extra power of $\mathcal{M}$.}
\end{figure}

The total tree-level squared amplitudes in the $S,T,U$ channels is evaluated to be 
\begin{equation}
\frac{128\pi^{2}\mathcal{M}^{2}}{g_{3}^{6}}|A|^{2}=-4|f^{abc}|^{2}\frac{s^{2}+st+t^{2}}{st(s+t)}
\end{equation} 
and is dressed by non-perturbative corrections. Inside the dark gluon condensate, axion emission will be kinematically possible if $m_{a}<2\Lambda_{D}$. This is compatible with QCD axion DM if $m_{a}\!\simeq \!10^{-1}\, \rm meV$. The evaporation rate is controlled by the scale $\mathcal{M}$. We can estimate a decay rate $\Gamma=\sigma n_{g} \!\sim\! n_{g}\mathcal{M}^{-2}$, with $n_g$ representing the density of the dark gluon condensate, which is roughly equal to $\Gamma \sim 10^{2}\Lambda_{D}^{3}\mathcal{M}^{-2}$ and then corresponds to $\tau\sim (\mathcal{M}/GeV)^{2}\, \rm Gyr$. In order not to have a complete evaporation of the condensate at the current cosmological epoch, we can set a limit $\mathcal{M}>40\, \rm GeV$, or same order of magnitude. A variation of $10\%$ in the DM density after a cosmological time of $10\, \rm Gyr$ can be visible if $\mathcal{M}\simeq 120\, \rm GeV$, or same order of magnitude. We also emphasize that the axions that are emitted have energy around $E\simeq 2\Lambda_{D}\simeq m_{a}$, therefore they are very slow and can easily be captured by axion cold DM condensate. Thus we may predict that DE density will be converted into cold DM density\footnote{It is worth to comment on some technical issues related to these rough estimates. First of all, for the dark gluon condensate at finite temperature there are thermal bath corrections to the axion emission processes. These corrections are negligible for $T\!<\!\!<\!\Lambda_{D}$, but will be relevant at earlier cosmological time, when $T\sim {\rm few}\, \Lambda_{D}\simeq 10^{-4}\, \rm eV$. These corrections can be computed by the general formalism of thermal QFT \cite{TFT}. Indeed, the thermal production rate of axions can be calculated from the imaginary part of its propagator $\Pi$, since $\gamma_{a}=\frac{d\Gamma}{dV}=-2\int \frac{d^{3}p}{(2\pi)^{3}2E} \Pi^{<}(p)$, in which $\Pi^{<}=f_{B}(E)\, {\rm Im} \Pi$. The thermal axions production rate can be obtained thermally averaging the scattering rate $gg\rightarrow g a$, or equivalently can be evaluated from the two-loops corrections to the axion propagator induced by gluons. Similar calculations for standard QCD axions were performed in \cite{Salvio:2013iaa}. Complete calculations of these contributions in our scenario are beyond the purpose of this letter. }.

\section{Conclusions}

\noindent 
In this paper we have investigated the possibility that dark energy is the vacuum energy of a YMC formed by a dark sector, and that at the same time a QCD invisible axion coupled to the latter sector provides a good candidate for cold DM. Both the hypotheses are highly motivated by string phenomenology. For instance, cancellation of the vacuum energy contributions from the other gauge sectors can be explained while resorting to $E8\times E8$ heterotic superstring theory with an asymmetric Higgs sector, or alternatively parallel intersecting D-brane worlds in open superstring theories\footnote{New intriguing implications in particle physics and cosmology of exotic stringy instantons have been studied within the framework of intersecting D-brane models. In particular exotic instantons can generate new effective operators not allowed at perturbative level, as an effective Majorana mass for the neutron, while the proton is not destabilized \cite{Addazi:2014ila,Addazi:2015rwa,Addazi:2015hka,Addazi:2015yna}.}.

Summarizing results of \cite{Dona:2015xia}, we have characterized the non-perturbative effective action and shown that, if it has a minimum in $\Theta$, a YMC forms and the model can actually work as DE at small $z$. If the effective action scales at least like the bare YM action for high-energy scale, at large $z$ it entails the EOS of radiation. Internal consistency requires that perturbative one-loop results must be still recovered in the appropriate asymptotic limit. In \cite{Dona:2015xia} these three general requirements have been checked successfully for a SU(2) YM-theory, by deploying non-perturbative FRG techniques. 

In the original part of this work, we have then sought a relation between YMC models of DE and DM models. In particular, we have shown that within the framework of a model $G_{SM}\times SU(2)_{D}\times U(1)_{PQ}$, QCD axions can be emitted by the YMC in a cosmological time. This causes a tiny conversion of a part of the DE density into cold DM density. This prediction can be tested and limited by next generations of experiments dedicated to accurate measures\footnote{Another possibility is the presence of a dark Born-Infeld condensate coupled with the axion field. In this case, the cosmological evolution of the condensate is expected to be different by the case we treated here \cite{Elizalde:2003ku, Addazi:2016oob}.} of DE. 

\section*{Acknowledgment}
\noindent
The authors gratefully acknowledge hospitality by LMU during the late stage of development of these ideas, and are grateful to Gia Dvali for extremely useful discussions.
Work of A.A. was supported in part by the MIUR research grant Theoretical Astroparticle Physics PRIN 2012CPPYP7 and by SdC Progetto speciale Multiasse La Societ\`a della Conoscenza in Abruzzo PO FSE Abruzzo 2007-2013. P.D. and A.M. wish to acknowledge support by the Shanghai Municipality, through the grant No. KBH1512299, and by Fudan University, through the grant No. JJH1512105.


\begin{thebibliography}{99}

\bibitem{Dona:2015xia} 
  P.~Don\`a, A.~Marcian\`o, Y.~Zhang and C.~Antolini,
  Phys. Rev. D {\bf 93} no.4, 043012 (2016) 
  [arXiv:1509.05824 [gr-qc]].

\bibitem{Riess:1998cb} 
  A.~G.~Riess {\it et al.}  
  Astron.\ J.\  {\bf 116}, 1009 (1998)
  [astro-ph/9805201].

\bibitem{Perlmutter:1998np} 
  S.~Perlmutter {\it et al.}  
  Astrophys.\ J.\  {\bf 517}, 565 (1999)
  [astro-ph/9812133].

\bibitem{Kowalski:2008ez} 
  M.~Kowalski {\it et al.}  
  Astrophys.\ J.\  {\bf 686}, 749 (2008)
  [arXiv:0804.4142 [astro-ph]].
  
\bibitem{Spergel:2003cb} 
  D.~N.~Spergel {\it et al.}  [WMAP Collaboration],
  Astrophys.\ J.\ Suppl.\  {\bf 148}, 175 (2003)
  [astro-ph/0302209].
  
  
\bibitem{Graham:2005xx} 
  A.~W.~Graham, D.~Merritt, B.~Moore, J.~Diemand and B.~Terzic,
  Astron.\ J.\  {\bf 132}, 2685 (2006)
  [astro-ph/0509417].
  
  
\bibitem{Spergel:2006hy} 
  D.~N.~Spergel {\it et al.}  [WMAP Collaboration],
  Astrophys.\ J.\ Suppl.\  {\bf 170}, 377 (2007)
  [astro-ph/0603449].
  
\bibitem{Komatsu:2008hk} 
  E.~Komatsu {\it et al.}  [WMAP Collaboration],
  Astrophys.\ J.\ Suppl.\  {\bf 180}, 330 (2009)
  [arXiv:0803.0547 [astro-ph]].
  
\bibitem{Cole:2005sx} 
  S.~Cole {\it et al.}  [2dFGRS Collaboration],
  Mon.\ Not.\ Roy.\ Astron.\ Soc.\  {\bf 362}, 505 (2005)
  [astro-ph/0501174].
  
\bibitem{Tegmark:2006az} 
  M.~Tegmark {\it et al.}  [SDSS Collaboration],
  Phys.\ Rev.\ D {\bf 74}, 123507 (2006)
  [astro-ph/0608632].

\bibitem{Capozziello:2003tk} 
  S.~Capozziello, S.~Carloni and A.~Troisi,
  Recent Res.\ Dev.\ Astron.\ Astrophys.\  {\bf 1}, 625 (2003)
  [astro-ph/0303041].

\bibitem{Carroll:2003wy} 
  S.~M.~Carroll, V.~Duvvuri, M.~Trodden and M.~S.~Turner,
  Phys.\ Rev.\ D {\bf 70}, 043528 (2004)
  [astro-ph/0306438].

\bibitem{AT} L.~Amendola and S.~Tsujikawa, {\it Dark Energy: Theory and Observations} (Cambridge University Press, Cambridge, UK, 2010).


\bibitem{YZ1} Y.~Zhang, {\it ``Inflation with quantum Yang-Mills condensate''}, Phys. Lett. B {\bf 340} (1994) 18-22. 

\bibitem{YZ2} T.~Y.~Xia and Y.~Zhang, {\it ``2-loop quantum Yang-Mills condensate as dark energy'',} Phys. Lett. B {\bf 656} (2007) 19-24.

\bibitem{YZ3} S.~Wang, Y.~Zhang and T.~Y.~Xia, {\it ``The three-loop Yang Mills condensate dark energy model and its cosmological constraints'',} JCAP {\bf 10} (2008) 037.

\bibitem{YZr1}
H. Pagels and E. Tomboulis, Nucl. Phys. B 143, (1978) 485

\bibitem{YZr2}
S.~Adler, Phys. Rev. D {\bf 23} (1981) 2905; S.~Adler, Nucl. Phys B {\bf 217} (1983) 3881; S.~Adler and T.~Piran, Rev. Mod. Phys. {\bf 56} (1984) 1; S.~Adler and T.~Piran, Phys. Lett. B {\bf 113} (1982) 405.

\bibitem{YZr3}
S.~G.~Matinyan and G.~K.~Savvidy, Nucl. Phys. B {\bf 134} (1978) 539.

\bibitem{YZr4} C.~N.~Yang, Phys Rev. Lett. {\bf 33} (1974) 445.


\bibitem{YZr5} W.~Zhao and Y.~Zhang, Phys. Lett. B {\bf 640} (2006) 69.

\bibitem{YZr6} S.~W.~Hawking and G.~F.~R.~Ellis, {\it ``The Large Scale Structure of Spacetime''}, Cambridge Univ. Press, 1973.

\bibitem{YZr7} L.~Parker and Y.~Zhang, Phys. Rev. D {\bf 44} (1991) 2421.


\bibitem{Savvidy:1977as}
  G.~K.~Savvidy,
  Phys.\ Lett.\ B {\bf 71} (1977) 133.
  
  
\bibitem{Weinberg:1980gg} S.~Weinberg,
{\it  In *Hawking, S.W., Israel, W.: General Relativity*, 790-831}
(Cambridge University Press, Cambridge, 1980).


   \bibitem{Wetterich:1993yh} C.~Wetterich,
`
Phys.\ Lett.\ B {\bf 301}, 90 (1993).
\bibitem{Morris:1993qb}
 T.~R.~Morris,
 Int.\ J.\ Mod.\ Phys.\ A {\bf 9} (1994) 2411
 [hep-ph/9308265].
 
\bibitem{Papenbrock:1994kf} 
  T.~Papenbrock and C.~Wetterich,
  Z.\ Phys.\ C {\bf 65}, 519 (1995)
  [hep-th/9403164].


 \bibitem{Fischer:2002hn}
   C.~S.~Fischer and R.~Alkofer,
   Phys.\ Lett.\ B {\bf 536}, 177 (2002)
   [hep-ph/0202202]. 


\bibitem{Fischer:2006vf}
  C.~S.~Fischer and J.~M.~Pawlowski,
  Phys.\ Rev.\  D {\bf 75}, 025012 (2007)
  [arXiv:hep-th/0609009];
  Phys.\ Rev.\  D {\bf 80}, 025023 (2009)
  [arXiv:0903.2193 [hep-th]].




\bibitem{Fischer:2008uz}
  C.~S.~Fischer, A.~Maas and J.~M.~Pawlowski,
  Annals Phys.\  {\bf 324} (2009) 2408
  [arXiv:0810.1987 [hep-ph]].


\bibitem{FRG}
U.~Ellwanger, M.~Hirsch and A.~Weber,
%
Z.\ Phys.\ C {\bf 69} (1996) 687; \\
U.~Ellwanger, M.~Hirsch and A.~Weber,
Eur.\ Phys.\ J.\  C {\bf 1} (1998) 563;\\
B.~Bergerhoff and C.~Wetterich,
%
Phys.\ Rev.\ D {\bf 57} (1998) 1591; [hep-ph/9708425].\\
J.~Kato,
hep-th/0401068. 



\bibitem{Pawlowski:2003hq}
J.~M.~Pawlowski, D.~F.~Litim, S.~Nedelko and L.~von Smekal,
Phys.\ Rev.\ Lett.\  {\bf 93} (2004) 152002 [hep-th/0312324]; \\
J.~M.~Pawlowski, D.~F.~Litim, S.~Nedelko and L.~von Smekal,
AIP Conf.\ Proc.\  {\bf 756}, 278 (2005)
[arXiv:hep-th/0412326]. 
  
  
  
\bibitem{Reuter:1994zn}
  M.~Reuter and C.~Wetterich,
  hep-th/9411227.
  
\bibitem{Reuter:1997gx}
  M.~Reuter and C.~Wetterich,
  Phys.\ Rev.\ D {\bf 56} (1997) 7893
  [hep-th/9708051].

\bibitem{Gies:2002af}
  H.~Gies,
  Phys.\ Rev.\ D {\bf 66} (2002) 025006
  [hep-th/0202207].


\bibitem{Eichhorn:2010zc}
  A.~Eichhorn, H.~Gies and J.~M.~Pawlowski,
  Phys.\ Rev.\ D {\bf 83} (2011) 045014
   [Phys.\ Rev.\ D {\bf 83} (2011) 069903]
  [arXiv:1010.2153 [hep-ph]].
  
\bibitem{Codello:2015oqa}
  A.~Codello, R.~Percacci, L.~Rachwał and A.~Tonero,
  Eur.\ Phys.\ J.\ C {\bf 76} (2016) no.4,  226
  doi:10.1140/epjc/s10052-016-4063-3
  [arXiv:1505.03119 [hep-th]].
  
 \bibitem{GC1}
  Y. A. Simonov, JETP Lett. {\bf 55}, 627 (1992).
  
  \bibitem{GC2}
  A. Peshier, B. Kampfer, O. P. Pavlenko, and G. Soff, Phys. Lett. B337, 235 (1994).
  
  \bibitem{GC3}
  M. I. Gorenstein and S.-N. Yang, Phys. Rev. D52, 5206 (1995).
  

\bibitem{Drago:2001gd}
  A.~Drago, M.~Gibilisco and C.~Ratti,
  Nucl.\ Phys.\ A {\bf 742} (2004) 165
  doi:10.1016/j.nuclphysa.2004.06.017
  [hep-ph/0112282].
  
\bibitem{Kolb:1990vq} 
  E.~W.~Kolb and M.~S.~Turner,
  Front.\ Phys.\  {\bf 69}, 1 (1990).

\bibitem{evaporation}  
J.~Preskill, M.~B.~Wise and F.~Wilczek, Phys. Lett. B {\bf 120}, 127 (1983); L.~F.~Abbott and P.~Sikivie, Phys. Lett. B {\bf 120}, 133 (1983); M.~Dine and W.~Fischler, Phys. Lett. B {\bf 120}, 137 (1983); P.~Sikivie, Lect. Notes Phys. {\bf 741}, 19 (2008).
  
  \bibitem{Graham:2015cka}
  P.~W.~Graham, D.~E.~Kaplan and S.~Rajendran,
  Phys.\ Rev.\ Lett.\  {\bf 115} (2015) 22,  221801
  doi:10.1103/PhysRevLett.115.221801
  [arXiv:1504.07551 [hep-ph]].
  
  \bibitem{PDG}
G.G. Raffelt and L.J. Rosenberg, Particle Data Group 2012, 
http://pdg.web.cern.ch/pdg/2013/reviews/rpp2012-rev-axions.pdf. 

\bibitem{Ejlli:2013uda}
  D.~Ejlli and A.~D.~Dolgov,
  Phys.\ Rev.\ D {\bf 90} (2014) 063514
  doi:10.1103/PhysRevD.90.063514
  [arXiv:1312.3558 [hep-ph]].
  
  \bibitem{TFT}
  M. Le Bellac, Thermal Field Theory, Cambridge University Press (2000), ISBN 0521654777.
  
\bibitem{Salvio:2013iaa}
  A.Salvio, A.Strumia and W.Xue,
  JCAP {\bf 1401} (2014) 011
  doi:10.1088/1475-7516/2014/01/011
  [arXiv:1310.6982 [hep-ph]].
  
\bibitem{Addazi:2014ila}
  A.Addazi and M.Bianchi,
  JHEP {\bf 1412} (2014) 089
  doi:10.1007/JHEP12(2014)089
  [arXiv:1407.2897 [hep-ph]].
  

  
\bibitem{Addazi:2015rwa}
  A.Addazi and M.Bianchi,
  JHEP {\bf 1507} (2015) 144
  doi:10.1007/JHEP07(2015)144
  [arXiv:1502.01531 [hep-ph]].
  
\bibitem{Addazi:2015hka}
  A.Addazi and M.Bianchi,
  JHEP {\bf 1506} (2015) 012
  doi:10.1007/JHEP06(2015)012
  [arXiv:1502.08041 [hep-ph]].
  

    
\bibitem{Addazi:2015yna}
  A.~Addazi, M.~Bianchi and G.~Ricciardi,
  JHEP {\bf 1602} (2016) 035
  doi:10.1007/JHEP02(2016)035
  [arXiv:1510.00243 [hep-ph]].
  
  
\bibitem{Elizalde:2003ku}
  E.~Elizalde, J.~E.~Lidsey, S.~Nojiri and S.~D.~Odintsov,
  Phys.\ Lett.\ B {\bf 574} (2003) 1
  doi:10.1016/j.physletb.2003.08.074
  [hep-th/0307177].
  
  
\bibitem{Addazi:2016oob} 
  A.~Addazi, S.~Capozziello and S.~Odintsov,
  Phys.\ Lett.\ B {\bf 760}, 611 (2016)
  doi:10.1016/j.physletb.2016.07.047
  [arXiv:1607.05706 [gr-qc]].

\end{thebibliography}
\end{document}